\documentclass{JINST}

\usepackage[T1]{fontenc} 
\usepackage{ae}
\usepackage{float} 
\usepackage{graphicx} 
\usepackage{amssymb} 
\usepackage{times} 
\usepackage{mathptmx} 
\usepackage{lineno} 

\title{Radio-assay of Titanium samples for the LUX Experiment}

\author{D.\,S. Akerib$^b$, X. Bai$^g$, S. Bedikian$^o$, E. Bernard$^o$, A. Bernstein$^e$, A. Bradley$^b$, S.\,B. Cahn$^o$, M.\,C. Carmona-Benitez$^b$, D. Carr$^e$, J.\,J. Chapman$^a$, Y-D. Chan$^d$, K. Clark$^b$, T. Classen$^j$, T.Coffey$^b$, S. Dazeley$^e$, L. de\,Viveiros$^a$, M. Dragowsky$^b$, E. Druszkiewicz$^m$, C.\,H. Faham$^a$, S. Fiorucci$^a$, R.\,J. Gaitskell$^a$, K.\,R. Gibson$^b$, C. Hall$^l$, M. Hanhardt$^g$, B. Holbrook$^j$, M. Ihm$^i$, R.\,G. Jacobsen$^i$, L. Kastens$^o$, K. Kazkaz$^e$, R. Lander$^j$,  N. Larsen$^o$, C. Lee$^b$, D. Leonard$^l$, K. Lesko$^d$, A. Lyashenko$^o$, D.\,C. Malling$^a$, R. Mannino$^h$, D. McKinsey$^o$, D. Mei$^n$, J. Mock$^j$, M. Morii$^c$, H. Nelson$^k$, J.\,A. Nikkel$^o$, M. Pangilinan$^a$,  P.\,D. Parker$^o$, P. Phelps$^b$, T. Shutt$^b$, W. Skulski$^m$, P. Sorensen$^e$, J. Spaans$^n$, T. Stiegler$^h$, R. Svoboda$^j$, A. Smith$^d$, M. Sweany$^j$, M. Szydagis$^j$, J. Thomson$^j$, 
M. Tripathi$^j$\thanks{Corresponding author.},
 J.\,R. Verbus$^a$, N. Walsh$^j$, R. Webb$^h$, J.\,T. White$^h$, M. Wlasenko$^c$, F.\,L.\,H. Wolfs$^m$, M. Woods$^j$, S. Uvarov$^j$, C. Zhang$^n$ \\
\llap{$^a$}Brown University, Dept. of Physics, 182 Hope St., Providence, RI 02912 \\
\llap{$^b$}Case Western Reserve University, Dept. of Physics, 10900 Euclid Ave, Cleveland, OH 44106 \\
\llap{$^c$}Harvard University, Dept. of Physics, 17 Oxford St., Cambridge, MA 02138 \\
\llap{$^d$}Lawrence Berkeley National Laboratory, 1 Cyclotron Rd., Berkeley, CA 94720 \\
\llap{$^e$}Lawrence Livermore National Laboratory, 7000 East Ave., Livermore, CA 94551 \\
\llap{$^f$}Moscow Engineering Physics Institute, 31 Kashirskoe shosse, Moscow 115409 \\
\llap{$^g$}South Dakota School of Mines and Technology, 501 East St Joseph St., Rapid City, SD 57701 \\
\llap{$^h$}Texas A \& M University, Dept. of Physics, College Station, TX 77843 \\
\llap{$^i$}University of California Berkeley, Dept. of Physics, Berkeley, CA 94720-7300 \\
\llap{$^j$}University of California Davis, Dept. of Physics, One Shields Ave., Davis, CA 95616 \\
\llap{$^k$}University of California Santa Barbara, Dept. of Physics, Santa Barbara, CA 95616 \\
\llap{$^l$}University of Maryland, Dept. of Physics, College Park, MD 20742 \\
\llap{$^m$}University of Rochester, Dept. of Physics and Astronomy, Rochester, NY 14627 \\
\llap{$^n$}University of South Dakota, Dept. of Physics, 414E Clark St., Vermillion, SD 57069 \\
\llap{$^o$}Yale University, Dept. of Physics, 217 Prospect St., New Haven, CT 06511\\
 E-mail: \email{mani@physics.ucdavis.edu}}

\abstract{
We report on the screening of samples of titanium metal for their radio-purity. The screening process described in this work led to the selection of  materials used in the  construction of the cryostats for the Large Underground
Xenon (LUX) dark matter experiment \cite{LUX-NIM}.  Our measurements establish titanium as a highly desirable material for low background experiments searching for rare events.
The sample with the lowest total long-lived activity was measured to contain 
 <0.25~mBq/kg of
$^{238}$U, <0.2~mBq/kg of $^{232}$Th, and <1.2~mBq/kg of $^{40}$K. Measurements of  several  samples also indicated the presence of short-lived (84 day half life) $^{46}$Sc, likely produced cosmogenically via muon initiated (n,p) reactions.}

\keywords{Materials for gaseous detectors, Materials for solid-state detectors, Noble-liquid detectors, Gamma detectors}

\begin{document}

\section{Introduction}

Experiments searching for rare events have to suppress background counts from natural radioactivity, which necessitates the minimization of radio-impurities embedded in the apparatus itself.  This is achieved by careful screening of construction
materials for the presence of long-lived radio-isotopes, such as $^{238}$U, $^{232}$Th and $^{40}$K.  The traditional method for this screening process consists of direct gamma-ray counting at low background facilities, located in laboratories with massive over-burden of rock.  

The LUX experiment, which aims to directly detect dark matter, is a two-phase xenon time-projection chamber, containing approximately 350 kg of xenon \cite{LUX-NIM}.  In order to sustain such a large mass of xenon in the liquid phase, two concentrically nested cryogenic vessels are employed, with a high vacuum maintained between the inner and outer chambers.  The combined mass of these titanium vessels is 230 kg. This article addresses the issue of the choice of material for these cryostats, describes the screening procedure, and presents the results of our measurements of the radio-purity for several titanium samples.

Traditionally, cryostats in large scale physics experiments have been constructed using stainless steel, due to its high tensile strength and relatively low cost; for situations in which radioactivity was critical, copper was preferred \cite{ZEPLIN}\cite{XENON100}. Titanium has not been used as the material of choice in any large dark matter experiment to date. The advantages of titanium over copper lie in its superior tensile strength, a somewhat lower cost, and the fact that it can be welded, as opposed to brazing for copper, which can leave micro-holes and allow radioactive contaminants such as radon to enter the experiment.  Certain types of stainless steel have been measured to have very low levels of radio-impurities \cite{Steel-purity} but still quite high compared to copper \cite{Cu-purity}. Hence, the choice between titanium and copper for the LUX cryostats rested on the radio-assays that are described here.

Titanium metal is available in several grades of chemical purity: the American Society for Testing and Materials (ASTM) lists 39 distinct grades \cite{ASTM-titanium}, with iron and oxygen content as the most important distinguishing feature.  The American Society of Mechanical Engineers (ASME) has its own nomenclature for categories of titanium metal. Addition of certain impurities, such as oxygen in the form of scrap titanium dioxide, improves the tensile strength of the alloy. Hence, our choice of material consisted of a compromise, because introduction of scrap materials in the processing of a metal carries with it the danger of parasitically introducing long-lived radio-isotopes.  We evaluated two grades of titanium in this study, colloquially labeled by vendors in terms of ``chemical purity" as CP-1 and CP-2. In this paper, CP-1 refers  to ASTM/ASME B-265 Grade 1 and CP-2 refers to ASTM B-265 Grade 2.  

Tables \ref{tab:thermal} and \ref{tab:npactivation} show the isotopes of titanium and their cross sections for thermal and fast neutron activation, respectively.  The only  isotope of concern is the moderately long-lived $^{46}$Sc, which is activated in an (n,p) interaction.  This process can be initiated by a cosmic ray muon interacting in the environment and producing neutrons via spallation.

Based on estimations of tolerable background rates in LUX, we established a conservative upper limit of 1 mBq/kg for a sample to be considered acceptable.  In order to ensure the lowest possible background from radio-isotopes in the cryostat material, several samples of Ti were evaluated in 2008-09.  In this article we present the results from this screening process and some follow-up measurements made in 2010-11. The work reported here was carried out at the Low Background Facility at  Berkeley Lab, the Oroville Counting Facility  in northern California, and the SOLO facility, located at the Soudan mine in northern Minnesota.

\begin{table}[p]
\caption{A table of stable isotopes of titanium and their activation cross sections when exposed to thermal neutrons.}
\label{tab:thermal}
\vskip 8pt
\begin{centering}
\begin{tabular}{| l | | c | c | c | c |}
\hline
\textbf{Isotope}  &  \textbf{Abundance} &    \textbf{Cross Section} &  \textbf{ Product} &   \textbf{Half-life}\\
&             \textbf{($\%$) } &    \textbf{(barns)} &         &             \textbf{(min)}\\
\hline
\hline
$^{46}$Ti & 8.0 & 0.59 & $^{47}$Ti  & stable\\
$^{47}$Ti  & 7.3 & 1.7 & $^{48}$Ti  & stable\\
$^{48}$Ti  & 73.8 & 7.84 & $^{49}$Ti & stable\\
$^{49}$Ti  & 5.5 & 2.2 & $^{50}$Ti  & stable\\
$^{50}$Ti  & 5.4 & 0.179 & $^{51}$Ti  & 5.77\\ 
\hline
\end{tabular}
\par\end{centering}
\end{table}

\begin{table}[p]
\caption{A table of stable isotopes of titanium and their activation cross sections via (n,p) reactions at a neutron energy of 5 MeV.}
\label{tab:npactivation}
\vskip 8pt
\begin{centering}
\begin{tabular}{| l | | c | c | c | c |}
\hline
\textbf{Isotope}  &  \textbf{Abundance} &    \textbf{Cross Section} &  \textbf{ Product} &   \textbf{Half-life}\\
&             \textbf{($\%$) } &    \textbf{at 5 MeV (mb)} &         &             \textbf{(hrs)}\\
\hline
\hline
$^{46}$Ti & 8.0  & 90.47 & $^{46}$Sc  & 2010.72 \\
$^{47}$Ti &7.3 & 61.40& $^{47}$Sc& 80.38 \\
$^{48}$Ti &73.8 & 0.14 & $^{48}$Sc & 43.67 \\
$^{49}$Ti &5.5  & 5.11 & $^{49}$Sc& 0.95 \\
\hline
\end{tabular}
\par\end{centering}
\end{table}

\section{Method}
\label{sec:method}

All the measurements reported in this paper were accomplished through the use of high-resolution gamma ray spectroscopy on samples by direct counting  with high-purity germanium (HPGe) detectors located in underground low-background facilities.  The detectors used for these measurements, and their Rn-purged shield enclosures, were constructed of low-background materials.  Multi-kilogram samples, ranging in weight from 1.5 to  8 kg, were analyzed to achieve the reported results.

We focussed on four benchmark isotopes in our screening of Ti metal samples:  $^{238}$U, $^{232}$Th, $^{40}$K, and $^{46}$Sc. $^{238}$U and $^{232}$Th are of particular concern to dark matter experiments for several reasons: a) each isotope decays through a chain in which a large number of high-energy gamma-rays are emitted, b) there is the possibility of generating neutrons via ($\alpha$,n) reactions in the surrounding material, and c) there is a small amount of direct generation of neutrons by spontaneous fission in the case of $^{238}$U. $^{46}$Sc is generated from cosmic ray interactions within the Ti, as described in Sec. \ref{sec:scandium}. $^{40}$K is a common contaminant in materials, which emits a 1461~keV gamma ray in its decay to $^{40}$Ar.

Individual isotope abundances are identified by their respective gamma-ray energies. $^{46}$Sc is identified by two $\sim$100\% branching ratio gamma-ray lines at 889 and 1120~keV.  $^{238}$U and $^{232}$Th cannot be directly identified by gamma-ray lines from decays of $^{238}$U and $^{232}$Th nuclei, as gamma-rays from these decays have very low branching ratios and energies. However, decays from isotopes later in the $^{238}$U and $^{232}$Th decay chains exhibit much higher detectability, and these gamma-rays are typically used to infer the concentration of $^{238}$U and $^{232}$Th. The strongest lines for identification of $^{238}$U come from $^{214}$Pb and $^{214}$Bi, and from $^{208}$Tl in the case of $^{232}$Th. $^{40}$K is identified by a single 1461~keV line.

We paid careful attention to the possibility that disequilibria may have been created in the U-series and/or the Th-series as a consequence of chemical processing of the titanium metal.   The most important causes for disequilibrium are differences between U/Th and Ra chemistry. Chemical processing during manufacturing can compromise the inference of $^{238}$U and $^{232}$Th from the late-chain isotopes mentioned above. For instance, a common breakage of equilibrium occurs in the $^{238}$U chain for the isotope $^{226}$Ra, which is soluble in water and can potentially be introduced or removed in large quantities during certain treatment processes, as can gaseous $^{222}$Rn. Because of this,  measurements of the $^{235}$U decay chain, even though they are less sensitive, are sometimes included as a direct measurement of ``early-chain" activity and the $^{238}$U content is deduced. The signatures include the 93~keV line from $^{234}$Th, the 186~keV line from $^{235}$U, and the 1001~keV line from $^{234}$Pa.  Note that the majority of gamma-rays contributing significantly to electromagnetic backgrounds in the detector are generated from late-chain decays. Hence, measurements of early-chain concentrations tend to result in very weak upper limits, and direct measurement of late-chain activity is typically considered to be of greater importance for background estimation.

\subsection{Facilities}

The Low Background Facility (LBF) at LBNL was established by A.R. Smith in the late 1950s for quantifying the stray radiation field of the newly operational Bevatron. Since the early 1980s, the LBF has been increasingly utilized to select low-activity construction materials for use in experiments searching for rare events. It features a 4$\pi$-shielded room with concrete walls (4-6 ft thick) made from about 500 tons of selected low-radioactivity serpentine rock (Mg$_6$Si$_{10}$OH$_8$) concrete.  The main detector, a 115\% N-type Low Background HPGe Detector, is housed in an outer Pb and an inner Cu (OFHC) shielding layer. The LBF is used as the first step in the screening process. If a positive measurement is made at  LBF, the screening process is over. If only an upper limit is obtained at LBF, the sample is sent to an offsite facility located under the Oroville dam in northern California.

The Oroville facility, which has 450 meters water equivalent (mwe) of shielding, provides a much higher sensitivity due to lower levels of backgrounds. An 85\% P-Type Ultra-low Background HPGe detector is housed within low-activity Pb and Cu shields and continuously flushed with nitrogen. It has now lived underground for about 8 years and has routinely provided sensitivities at levels of $<$1~mBq/kg $^{238}$U/$^{232}$Th for late-chain measurements.  The group at LBNL has established this screening procedure over more than 4 decades, and their results are considered an industry standard.  

Data were also taken at the SOLO facility \cite{SOLO} located at the Soudan Underground Laboratory, which boasts a 2200~mwe of overhead shielding.  A 0.6~kg HPGe detector is housed in a lead shield >30~cm thick, with the inner 5~cm comprised of ancient lead with $^{210}$Pb activity measured below 50~mBq/kg. The detector offers sensitivities at the level of $\sim$1~mBq/kg $^{238}$U/$^{232}$Th for late-chain measurements.  The SOLO facility is operated by the LUX collaboration and  is available for our use at all times. However, it  was in the process of being commissioned at the time of this study and was mostly used for the Sc studies reported in this paper.

\section{Results of Radio-assays}

Nine different samples of Ti (seven samples of CP-1and two samples of CP-2), obtained from various distributors, were counted at the three different facilities between February 2008 and August 2011. Samples Ti1 - Ti7 are plate samples, while Ti8 is a sheet, and Ti9 is a sample of Ti welding wire. Sample descriptions are listed in Table \ref{tab:Sample-info}. 

Counting results for CP-1 samples are listed in Table \ref{tab:CP-1results}.  All quoted errors are at $\pm$1$\sigma$, while upper limits are at 90\% confidence level (CL).  The (e) and (l) labels refer to early- and late-chain measurements, as described in Sec \ref{sec:method}. Ti1, Ti4, Ti5, and Ti6 are plate samples of similar thicknesses and consistently show only upper limits for radioactive contaminations, except in the case of Ti6, for which an early-chain estimate of  6.2$\pm$1.2~mBq/kg was made from measuring the 186~keV line from $^{235}$U~ decays.  This sample has the lowest measured $^{238}$U upper limit at <0.19~mBq/kg, and was ultimately used for the construction of the body of the LUX cryostats. While samples Ti1, Ti4, and Ti5 did not have a positive count for early-chain $^{238}$U, the low late-chain counts of Ti6 make it the superior material.

Ti7 is a thick plate sample which was used to fabricate the sealing flanges and end-caps for the LUX cryostats. It has the lowest measured $^{232}$Th upper limit at <0.2~mBq/kg and $^{238}$U upper limit at <0.25~mBq/kg.  Sample Ti8 was used in small quantities for fabricating peripheral parts.  Hence, even though it shows positive
measurements for late-chain decays, the total mass used in LUX was negligible compared to the bulk materials (Ti6 and Ti7).
Ti9, which is the welding wire used in construction, also has a positive
signature for both early- and late-chain decays. The Ti9 late-chain
signature is more than twice that measured for Ti8, more than twice
the highest upper limit (made for sample Ti4), and $\times$28 higher
than the lowest upper limit placed.  Again, negligible quantities of this material were employed.

Results for CP-2 are listed in Table \ref{tab:CP-2results}. Sample Ti3 was measured with a large positive
$^{238}$U contamination, at 37$\pm$12 mBq/kg. We were not able to place any strong limit on $^{238}$U contamination of sample Ti2,
as the baseline activity was significantly raised from Compton scattering of 2.6~MeV gamma-ray
associated with the highly abundant $^{232}$Th in this sample.

Comparison of counting results for CP-1 and CP-2 reveals that CP-1 has a much lower average $^{232}$Th content than CP-2.  A difference in $^{238}$U contamination is also seen between CP-1 and CP-2. 
Measurements for CP-1 $^{238}$U contamination
yield only upper limits for four out of seven samples, with a mean positive
measurement of 9.9$\pm$1.8~mBq/kg (early) for the remaining three. Late-chain activities for CP-1 are significantly lower, with only samples Ti8 and Ti9
indicating positive detection at an average of 2.1~mBq/kg. 

No positive signatures of $^{40}$K were identified, except in the case of the wire sample Ti9. Given the positive identification of $^{238}$U / $^{232}$Th at levels higher than typical for other CP-1 materials (particularly in the case of $^{232}$Th), as well as the presence of $^{40}$K, it is likely that this sample was contaminated during chemical processing. It should be noted that the wires were not deliberately thoriated, as is the practice for welding applications.  

\begin{table}[p]
\caption{Descriptions of all Ti samples used in the radio-assay for this work. Sample IDs are matched to counting results given in the tables below.}
\label{tab:Sample-info}
\vskip 8pt
\begin{centering}
\begin{tabular}{| l || c | c | c | c  | c |}
\hline 
\textbf{ID}   & \textbf{Grade}    & \textbf{Type}      & \textbf{Dimensions}  & \textbf{Quantity} & \textbf{Total Mass} \\
 &    &    & \textbf{(inches)}     &      & \textbf{(kg)}  \\
\hline
\hline 
Ti1                & CP-1        &Plate	& 2.5$\times$6$\times$0.375        & 4        & 1.873 \\
Ti2                & CP-2       &Plate	& 4$\times$6$\times$0.188        & 20        & 7.55  \\
Ti3                & CP-2        &Plate	& 1.25$\times$6$\times$0.358         & 8        & 1.546 \\
Ti4                & CP-1        &Plate	& 4$\times$6$\times$0.188        & 18        & 7.34 \\
Ti5                & CP-1        &Plate	& 4$\times$6$\times$0.188        & 16        & 6.51\\
Ti6                & CP-1        &Plate	& 4$\times$6$\times$0.188        & 20        & 7.978\\
Ti6-A            & CP-1        &Plate	& 4$\times$6$\times$0.188        & 9          & 3.586\\
Ti7                & CP-1        &Plate	& 1.875$\times$6$\times$1        & 8  &  7.201\\
Ti8                & CP-1        &Sheet	& 4$\times$6$\times$0.0625        & 40        & 4.399\\
Ti9                & CP-1        &Wire	& 0.0625 dia. x 6 long        & 780 & 4.189 \\
\hline
\end{tabular}
\par\end{centering}
\end{table}

\begin{table}[p]
\caption{Radioactivity assay results for 7 different CP-1 Ti samples. The early- (e) and late-chain (l) measurements are reported separately as applicable. All reported errors are at $\pm$1$\sigma$ level for the positive measurements. All upper limits are at 90\% CL.  The sample Ti6-A had a lower mass than sample Ti6, as reported in Table 3.  The materials represented by samples Ti6 and Ti7  were used as the bulk materials for the construction of LUX cryostats.  Materials represented by samples Ti8 and Ti9 were used in small quantities for fabricating peripheral parts and for welding, respectively.}
\label{tab:CP-1results}
\vskip 8pt
\begin{centering}
\begin{tabular}{| c || l | c | c | c | c | c |}
\hline 
\textbf{ID}     & \textbf{Facility}   & \textbf{Counting}  & \textbf{$^{238}$U~}   & \textbf{$^{232}$Th~} & \textbf{$^{40}$K~}  & \textbf{$^{46}$Sc~}  \\
                      &                              & \textbf{Period}       & \textbf{(mBq/kg)}    & \textbf{(mBq/kg)} & \textbf{(mBq/kg)}   & \textbf{(mBq/kg)}          \\
\hline
\hline 
Ti1                & Oroville               & Feb 2008               & <2.5                & <1.6             & <6.2                             & 4.8  \\
\hline
Ti4                & LBF                      &  Jun 2008              & <3.8                & <2.8             & <9.3                             & 3.0  \\
                      & LBF                      &  Oct 2010               & <6.3 (e)          & <1.2             & <9.3                             &         \\
                      &                              &                                 & <1.3 (l)          &                        &                                      &         \\
\hline
Ti5                & SOLO                  &   2008                      & <2.5 (l)          & <0.8 (l)        &                                      & <0.35 \\
\hline
Ti6               & Oroville               & Sep 2008               & 6.2$\pm$1.2 (e)  & <0.24     & <0.93                            & 23$\pm$1 \\
                      &                              &                                 & <0.19 (l)         &                     &                                     &  \\
Ti6-A            & LBF                      & Aug 2011               & <11.3 (e)       & <1.6           & <9.3                             &  \\
                      &                              &                                 & <0.25 (l)         &                     &                                     &  \\
\hline
Ti7                & Oroville               & Jan 2009               & <0.25             & <0.2            & <1.2                            & 2.5  \\
\hline
Ti8                & Oroville               & Sep 2009              & 10$\pm$3.8 (e) & <0.8 (e)    & <1.9                        & 6$\pm$1\\
                     &                               &                                 & 2.4$\pm$0.5 (l) & 2.8$\pm$0.4 (l)   &                 &   \\
\hline
Ti9                & Oroville               & Sep 2009              & 15$\pm$3.8 (e)   & 8.1$\pm$1.2 (e)    & 7.4$\pm$1.9  & \\
                     &                               &                                & 1.9$\pm$0.4 (l)  & 5.7$\pm$0.8 (l)    &                           & \\
\hline
\end{tabular}
\par\end{centering}
\end{table}

\begin{table}[p]
\caption{Radioactivity assay results for two different CP-2 Ti samples. The early- (e) and late-chain (l) measurements are reported separately as applicable. All reported errors are at $\pm$1$\sigma$ level for the positive measurements. All upper limits are at 90\% CL.  Materials represented by these samples were not used in the construction of the LUX detector.}
\label{tab:CP-2results}
\vskip 8pt
\begin{centering}
\begin{tabular}{| c || l | c | c | c | c | c |}
\hline 
\textbf{ID}     & \textbf{Facility}   & \textbf{Counting}  & \textbf{$^{238}$U~}   & \textbf{$^{232}$Th~} & \textbf{$^{40}$K~}  & \textbf{$^{46}$Sc~}  \\
                      &                              & \textbf{Period}       & \textbf{(mBq/kg)}    & \textbf{(mBq/kg)} & \textbf{(mBq/kg)}   & \textbf{(mBq/kg)}          \\
\hline
\hline 
Ti2                & SOLO          &  2008             & <140 (l)                 & 70$\pm$5 (l)     & <9.3                       &            \\
\hline
Ti3                & SOLO         &   2008              & <27 (l)  & <22 (l)&                      & <23   \\
&&&&&&\\
                  & Oroville  & Nov 2010                 & 37$\pm$12 (e)    & 5.7$\pm$2 (e) & <5                      &  \\
                     &                      &                  & 7.4$\pm$1.2 (l)   & 16$\pm$2 (l)    &                              &            \\
\hline
\end{tabular}
\par\end{centering}
\end{table}

\subsection{Measurements of Cosmogenic Activation}
\label{sec:scandium}
 
Cosmogenic activation, while possible at sea level, is enhanced during air shipment. The primary cosmogenic activation product in Ti is $^{46}$Sc, with a half-life of $\sim$84 days. It decays by emitting two $\sim$1 MeV gamma-rays (889 and 1120
keV), along with a beta with an endpoint energy of 356 keV. The cross section for the dominant
production via fast neutron activation, $^{46}\mbox{Ti}(\mbox{n},\mbox{p})$$^{46}$Sc, is $\sigma\sim0.25~\mbox{barn}$
for neutron energies in the range 5-20~MeV. Neutrons in this energy range are predominantly generated from muon spallation in nearby materials. 

Positive measurements of $^{46}$Sc were made for several of the Ti
samples, as shown in Tables \ref{tab:CP-1results} and \ref{tab:CP-2results}.  In some cases, upper limits were extracted, but not for each sample.  The measured activities range from below 0.19 mBq/kg to values as high as 23~mBq/kg (Ti6).  We believe that this variation is due to their history of transportation, which was not monitored prior to their arrival in our labs.  A subset of Ti6, listed as Ti6-A, was counted again after a gap of about 3 years and no peaks due to $^{46}$Sc were visible.  Similarly, Ti3 underwent a count after about a two-year gap, and no $^{46}$Sc lines were observed.  This leads us to conclude that cosmogenic $^{46}$Sc will not be an issue for titanium stored at locations near sea level (in this case, Davis, CA).  However, the surface lab at Lead, SD is situated about 1,500 m above sea level, and presents possible concerns for cosmogenic activity.

We did a more detailed study of $^{46}$Sc activation using sample Ti5.  It was initially counted
at SOLO after being underground for $\sim$300 days, yielding <0.35~mBq/kg of
$^{46}$Sc. It was then sent to Sanford Lab  for a period of
$\sim$6 months, after which it was shipped by ground back to SOLO
for counting. Counting after shipment yielded an activity of 4.4$\pm$0.3~mBq/kg,
averaged over a counting period of 14~days.  Ideally, we should expose a sample to cosmic rays for a period about 5 times larger than the half-life of the activation product, in order to achieve equilibrium levels.  However, this measurement provides us with a good estimate for the equilibrium level of $^{46}$Sc that can be expected in all titanium parts that were stored at the Sanford Lab for a period of 6 months or more.  

\section{Conclusions}

From the nine Ti samples screened in the LUX materials selection program, the CP-1 plate samples were found
to have a significantly reduced $^{232}$Th and $^{238}$U content compared to the CP-2 samples.  The CP-1 sheet and wire samples were not significantly different from CP-2 plate samples.  The bulk materials used in the construction of the LUX cryostat were measured to have total activities well below 1~mBq/kg, our established tolerance level.  Within our limited data set, it is difficult to draw definitive conclusions about radioactive properties of various types of titanium metal.  However, it is evident that ASTM/ASME B-265 Grade 1 (CP-1) titanium is a very promising material for low background experiments.  Prolonged exposure of titanium to cosmic rays at sea level is not of concern, but storage at high altitudes causes a build-up of $^{46}$Sc levels.

\acknowledgments{
The LUX collaboration would like to express its thanks to Howard Nicholson for suggesting that we pursue titanium. We gratefully acknowledge the help of Jim Beaty and Dave Saranen at Soudan Underground Laboratory for their help in counting the Ti samples with SOLO.  This work was partially supported by the U.S. Department of Energy (DOE) grants DE-FG02-08ER41549, DE-FG02-91ER40688, DE-FG02-95ER40917, DE-FG02-91ER40674, DE-FG02-11ER41738, DE-FG02-11ER41751, DE-AC52-07NA27344, the U.S. National Science Foundation grants PHY-0750671, PHY-0801536, PHY-1004661, PHY-1102470, and PHY-1003660, the Research Corporation grant RA0350, the Center for Ultra-low Background Experiments at DUSEL (CUBED), and the South Dakota School of Mines and Technology (SDSMT). We gratefully acknowledge the logistical and technical support and the access to laboratory infrastructure provided to us by the Sanford Underground Research Facility (SURF) and its personnel in Lead, SD.
}

\end{document}